\documentclass[11pt,epsf,letterpaper]{article}%
\usepackage{color}
\usepackage{amsmath}
\usepackage{amsfonts}
\usepackage{verbatim}
\usepackage{amssymb}
\usepackage{graphicx}
\usepackage{epstopdf}
\usepackage{mathrsfs}
\usepackage{epsfig}
\usepackage{cite}%
\providecommand{\U}[1]{\protect\rule{.1in}{.1in}}
\begin{document}

\title{Hairy Black Hole Stability in AdS, Quantum Mechanics on the Half-Line and Holography.}
\author{Andr\'{e}s Anabal\'{o}n$^{(1)}$, Dumitru Astefanesei$^{(2)}$ and Julio
Oliva$^{(3)}$\\\textit{$^{(1)}$Departamento de Ciencias, Facultad de
Artes Liberales y}\\\textit{Facultad de Ingenier\'{\i}a y Ciencias,
Universidad Adolfo Ib\'{a}\~{n}ez,}\\\textit{Av. Padre Hurtado 750,
Vi\~{n}a del Mar, Chile.}\\\textit{$^{(2)}$Instituto de F\'{\i}sica,
Pontificia Universidad Cat\'{o}lica de Valpara\'{\i}so,} \\\textit{
Casilla 4059, Valpara\'{\i}so, Chile.}\\\textit{$^{(3)}$
Departamento de F\'{\i}sica, Universidad de Concepci\'{\o}n, Casilla
160-C, Concepci´on, Chile}
\\andres.anabalon@uai.cl, dumitru.astefanesei@ucv.cl,
julio.oliva@uach.cl} \maketitle

\begin{abstract}
We consider the linear stability of $4$-dimensional hairy black holes with
mixed boundary conditions in Anti-de Sitter spacetime. We focus on the mass of
scalar fields around the maximally supersymmetric vacuum of the gauged
$\mathcal{N}=8$ supergravity in four dimensions, $m^{2}=-2l^{-2}$. It is shown
that the Schr\"{o}dinger operator on the half-line, governing the $S^{2}$,
$H^{2}$ or $\mathbb{R}^{2}$ invariant mode around the hairy black hole, allows
for non-trivial self-adjoint extensions and each of them corresponds to a
class of mixed boundary conditions in the gravitational theory. Discarding the
self-adjoint extensions with a negative mode impose a restriction on these
boundary conditions. The restriction is given in terms of an integral of the
potential in the Schr\"{o}dinger operator resembling the estimate of Simon for
Schr\"{o}dinger operators on the real line. In the context of AdS/CFT duality,
our result has a natural interpretation in terms of the field theory dual
effective potential.

\end{abstract}

\section{Introduction}

A form of understanding the asymptotically flat no-hair theorems is
that they connect the stability of the ground state and the
existence of hairy black holes. Namely, if the relevant
scalar-tensor theory has a stable Minkowski vacuum then it does not
admit hairy black hole configurations. Indeed, the Bekenstein
no-hair theorem shows that a convex scalar field potential implies
the non-existence of hairy black hole solutions
\cite{Bekenstein:1974sf}. The same has been shown to be true for
non-negative potentials \cite{Heusler:1992ss, Sudarsky:1995zg}.
Moreover, it is now clear that a necessary condition for the
existence of asymptotically flat hairy black hole solutions is a
scalar field potential with a negative region \cite{SUD2, AO,
Anabalon:2013qua, Cadoni:2015gfa}. Therefore, if the scalar field
potential has a extremum allowing for a Minkowski vacuum, where the
scalar field potential necessarily vanishes, a sufficiently large
perturbation will naturally explore the sector where the potential
is negative, destabilizing the Minkowski ground state. Hence, one
may wonder whether the hairy black hole itself is stable. However,
it happens that all the asymptotically flat hairy black holes
described in the literature are unstable under spherically symmetric
perturbations \cite{Harper:2003wt, H2, Dotti, Broni, Der}. On the
contrary, odd-parity perturbations are generically well behaved
around hairy black holes \cite{Anabalon:2014lea}. For a recent
review of the asymptotically flat case see \cite{Herdeiro:2015waa}.
Indeed, one is having in mind here a single, real scalar field.
Already the case of a complex scalar field is different. The black
holes no-hair theorems can be circumvented in this case because the
scalar field and the metric do not share the same symmetries (but
the energy momentum tensor and the metric do)
\cite{Herdeiro:2014goa, Herdeiro:2015gia}.

However, for a single and real scalar field, the picture is
completely different in asymptotically AdS spacetimes in the sense
that it is not necessary to have a positive scalar field potential
to have a stable AdS vacuum. As is well-known, scalar fields with
Dirichlet boundary conditions and masses above the
Breitenlohner-Freedman (BF) bound,
\begin{equation}
m_{BF}^{2}=-\frac{(D-1)^{2}}{4l^{2}}\text{ ,}%
\end{equation}
where $D$ is the spacetime dimension and $l$ is the AdS radius, do not
generate linear instabilities in global AdS spacetimes
\cite{Breitenlohner:1982bm, BF}.

When the squared scalar field mass, $m^{2}$, is above (or saturates) the BF
bound and strictly less than the unitarity bound,%
\begin{equation}
m_{BF}^{2}\leq m^{2}<m_{BF}^{2}+l^{-2}\text{ }, \label{window}%
\end{equation}
the Klein-Gordon operator has non-trivial self-adjoint extensions in
global AdS spacetime \cite{Ishibashi:2004wx} (for a pedagogical
discussion of self-adjoint extensions see \cite{Bonneau:1999zq}).
Each of these self-adjoint extensions defines a class of theories
and some of these theories are equivalently defined by the
requirement of the existence of a soliton with a given value of the
scalar field at the origin. Such constructions go by the name of
designer gravity \cite{Hertog:2004ns}. Furthermore, these
generalized boundary conditions are present for all examples of
hairy black holes known so far \cite{Torii:2001pg, Henneaux:2002wm,
Sudarsky:2002mk, Hertog:2004dr, MTZ, Kolyvaris:2009pc,
Correa:2010hf, Anabalon:2012sn, Anabalon:2012ta, Acena:2012mr,
Anabalon:2012dw, Aparicio:2012yq, Zhao:2013isa, Lu:2013ura,
Gonzalez:2013aca, Bueno:2013vua, Acena:2013jya, Anabalon:2013eaa,
Feng:2013tza, Zhang:2014sta, Gonzalez:2014tga, Zhang:2014dfa,
Faedo:2015jqa, Fan:2015tua, Fan:2015oca}.

There is a no-hair conjecture proposed by Hertog
\cite{Hertog:2006rr} that, modulo certain technical requirements,
captures the spirit of the asymptotically flat case. It states that,
if there exists a suitable superpotential that is globally defined,
the boundary conditions are AdS invariant, and there is a bound of
the energy\footnote{Mixed boundary conditions introduce subtleties
to the existence of a bound \cite{Faulkner:2010fh}, see
\cite{Boucher:1984yx,Townsend:1984iu}.}, then the theory does not
allows the existence of hairy black holes. Global AdS has vanishing
mass for all the boundary conditions, hence it should saturate the
bound and, if there is no hairy soliton, its non-perturbative
stability would be ensured. Within the conditions of the conjecture,
non-perturbative stability of the ground state imply the no-hair
condition.

In this paper we shall consider the less explored subject of hairy
black hole stability in asymptotically AdS spacetimes. The theory of
black hole stability was developed by Regge-Wheeler and Zerilli
\cite{Regge:1957td, Zerilli:1970se}. The higher dimensional
generalization was done by Ishibashi-Kodama \cite{Kodama:2003jz}.
The approach to the problem is to linearize the system around a
background and to classify the perturbations as representations of
the background isometries. The dynamics for each type of
perturbation decouples satisfying a wave equation with an effective
potential. It is straightforward to show that, in the frequency
domain, the modes satisfy a Schr\"{o}dinger equation. When the
spectrum is positive and the relevant domain of the operator is
$L^{2}\left(  {\mathbb{R}}\right)  $, it has a unique self-adjoint
extension, namely the Friedrichs extension, which implies that the
linearized evolution is well defined and the background is stable,
at least in the static case \cite{Ishibashi:2003ap}.

AdS is not globally hyperbolic, which means that to make sense of the
evolution it is necessary to provide the initial data and impose certain
boundary conditions. Namely, around a black hole background, the
Schr\"{o}dinger operator governing the dynamics acts on $L^{2}\left(  \left]
{\mathbb{-\infty}},0\right]  \right)  $. Hence, it has an infinite number of
self-adjoint extensions parameterized by a real number. For the case at hand,
and even if the effective potential is everywhere positive, the self-adjoint
extensions modify the spectrum of the Schr\"{o}dinger operator introducing
exactly one negative eigenvalue. As we shall see below, these non-trivial
self-adjoint extensions are associated with the existence of the two
normalizable modes in the window (\ref{window}). The main objective of this
paper is to establish which boundary conditions implies the existence of this
instability, thus providing a simple tool to analyze a complex problem. The
condition is sufficient and, if one would be interested in finding a stable
hairy black hole, it allows to reduce the number of cases relevant to study in
the first place.

Whenever the gravitational theory has a CFT dual and the boundary
conditions are AdS invariant, they correspond to multi-trace
deformations of the boundary CFT \cite{Witten:2001ua} --- for a nice
discussion with the detailed holographic dictionary, see
\cite{Papadimitriou:2007sj}. As formulated in \cite{Hertog:2004ns},
the non-perturbative stability of AdS corresponds to a convex
potential in the field theory with only one global minimum. If the
effective potential has several global minima, with zero energy,
they are identified with the existence of hairy solitons. A clear
holographic picture thus arises connecting stability of the
gravitational theory and of the field theory. Furthermore, the dual
interpretation of hairy black holes (and their stability) in terms
of a finite temperature version of the field theory effective
potential was first put forward and analyzed in
\cite{Hertog:2005hu}. Indeed, our condition for hairy black hole
instability seems to be suited for an interpretation along the lines
of \cite{Hertog:2004ns, Hertog:2005hu} and we comment on it. This
shows a clear connection between the mechanical stability of hairy
black holes and the study of the positive energy theorems. There has
been a lot a of work on proving positive energy theorems for certain
classes of designer gravity theories. The first theorem for bounded
effective potentials were proven in \cite{Hertog:2005hm} and this
was generalized to certain classes of effective potential that are
even unbounded below in \cite{Faulkner:2010fh}.

This kind of instability at the conformal mass has been already observed by
Martinez \cite{Martinez:1998db} in the particular case of the Martinez-Zanelli
hairy black hole \cite{Martinez:1996gn}, whose source is a conformally coupled
scalar field in three dimensions. The Schr\"{o}dinger operator governing the
evolution of the linearized dynamics has an everywhere non-negative effective
potential around the Martinez-Zanelli hairy black hole, however a normalizable
mode that is exponentially growing in time is shown to exist. Our analysis
sheds light on this counterintuitive result.

\section{Linear stability}

The theory of the linear spherically symmetric perturbations in the case of
gravity coupled to scalar fields dates back to \cite{Harper:2003wt}, more
recently it has been discussed in \cite{H2, Dotti, Broni, Der}. Here, we
provide a straightforward generalization to the hyperbolic and planar modes.
The action principle is given by%
\begin{equation}
S[g_{\mu\nu},\phi]=\int d^{4}x\sqrt{-g}\left[  \frac{R}{2\kappa}-\frac{1}%
{2}\left(  \partial\phi\right)  ^{2}-V(\phi)\right]  \text{ }.
\end{equation}
In this section $V(\phi)$ is an arbitrary function of $\phi$ with at least one
AdS vacuum. The field equations are%
\begin{equation}
E_{\mu\nu}=R_{\mu\nu}-\frac{1}{2}g_{\mu\nu}R-\kappa T_{\mu\nu}\text{ },
\label{EFEQ}%
\end{equation}
When the Einstein equations are satisfied, the scalar field equation holds as
a consequence of the conservation of the energy momentum tensor
\begin{equation}
T_{\mu\nu}=\partial_{\mu}\phi\partial_{\nu}\phi-g_{\mu\nu}\left[  \frac{1}%
{2}\left(  \partial\phi\right)  ^{2}+V(\phi)\right]  \text{ }.
\end{equation}
Let us consider the four dimensional time dependent metric%

\begin{equation}
ds^{2}=-\Delta_{1}(x,t)dt^{2}+\Delta_{2}(x,t)dx^{2}+C(x,t)d\Sigma_{k}\text{ ,}
\label{coor}%
\end{equation}
where $d\Sigma_{k}$ is the metric of a two dimensional space of constant
curvature, with Ricci scalar normalized to $2k,$ and the scalar field%
\begin{equation}
\phi=\phi(x,t)\text{ }.
\end{equation}
It is always possible to use the differmorphism invariance to set
$C(x,t)=C(x),$ where, for the sake of generality, we choose to let $C(x)$
arbitrary. We are interested in studying the linearized dynamics around a
background solution. Hence, we introduce the expansion
\begin{align}
\Delta_{1}(x,t)  &  =A(x)+\epsilon A_{1}(x,t)\text{ ,}\\
\Delta_{2}(x,t)  &  =B(x)+\epsilon B_{1}(x,t)\text{ ,}\\
\phi(x,t)  &  =\phi_{0}(x)+\epsilon\phi_{1}(x,t)\text{ ,}\\
V(\phi)  &  =V_{0}+\epsilon V_{1}\phi_{1}(x,t)\text{ ,}%
\end{align}
where%
\begin{equation}
V_{0}=V(\phi_{0})\qquad,\qquad V_{n}=\left.  \frac{d^{n}V}{d\phi^{n}%
}\right\vert _{\phi=\phi_{0}}\text{ .}%
\end{equation}
From now on, we shall assume that the Einstein field equations are solved by
the $\epsilon=0$ set of functions. Birkhoff's theorem ensures that all the
dynamics is driven by the scalar field. Indeed, it is possible to write the
metric perturbations in terms of the $\phi_{1}(x,t)$ by using the Einstein
field equations. Let us consider first:%
\begin{equation}
E_{tx}=\left(  \frac{C^{\prime}}{2BC}\dot{B}_{1}-\kappa\dot{\phi}_{1}\phi
_{0}^{\prime}\right)  \epsilon+O\left(  \epsilon^{2}\right)  \text{ ,}
\label{E1}%
\end{equation}
where $\dot{\phi}_{1}=\partial_{t}\phi_{1}$ and $\phi_{1}^{\prime}%
=\partial_{x}\phi_{1}$. Thus, modulo a trivial redefinition of the
perturbations it follows from (\ref{E1}) that%
\begin{equation}
B_{1}(x,t)=2\kappa\frac{CB}{C^{\prime}}\phi_{1}\phi_{0}^{\prime}\text{ ,}
\label{beta}%
\end{equation}
We replace (\ref{beta}) in the remaining Einstein equations, then
automatically follows that $E_{tt}=O(\epsilon^{2})$. The radial equation is
also a simple constraint%

\begin{equation}
E_{xx}=\left(  -\frac{1}{2}\frac{C^{\prime}A^{\prime}}{CA^{2}}A_{1}+\frac
{1}{2}\frac{C^{\prime}}{CA}A_{1}^{\prime}+\kappa\frac{B}{C^{\prime}}\left[
-2k\phi_{0}^{\prime}+V_{1}C^{\prime}+2\kappa CV_{0}\phi_{0}^{\prime}\right]
\phi_{1}-\kappa\phi_{0}^{\prime}\phi_{1}^{\prime}\right)  \epsilon+O\left(
\epsilon^{2}\right)  \text{ .} \label{E2}%
\end{equation}
Hence, it is possible to obtain $A_{1}^{\prime}$ from (\ref{E2}). When
$A_{1}^{\prime}$ and $B_{1}$ are replaced in the equation along the
coordinates of the sphere a messy expression is obtained. However, this
equation can be simplified by introducing a master variable
\begin{equation}
\psi(z,t)=\phi_{1}(x,t)C(x)^{1/2}\text{ ,}%
\end{equation}
where $z~$is the tortoise coordinate%
\begin{equation}
dz=\left(  \frac{B}{A}\right)  ^{1/2}dx\text{ .} \label{tor}%
\end{equation}
The master equation is
\begin{equation}
-\partial_{z}^{2}\psi+U\psi=-\partial_{t}^{2}\psi\text{ ,} \label{Schr}%
\end{equation}
with the effective potential%
\begin{equation}
\frac{U}{A}=4\kappa C\left[  \left(  \kappa V_{0}C-k\right)  \left(
\frac{d\phi_{0}}{dC}\right)  ^{2}+V_{1}\left(  \frac{d\phi_{0}}{dC}\right)
\right]  -\kappa V_{0}+V_{2}+\frac{k}{C}-\frac{1}{4B}\left(  \frac{C^{\prime}%
}{C}\right)  ^{2}\text{ }. \label{pot}%
\end{equation}

If $z$ takes its values in the whole real line and $U$ is non-negative, the
operator (\ref{Schr}) is essentially self-adjoint and its spectrum is positive
which implies that the background is mode stable under spherically symmetric
perturbations\footnote{Non-modal stability is harder to prove and just
recently has been confirmed for the Schwarzschild black hole
\cite{Dotti:2013uxa}.}. It is possible to extract some generic behaviour of
$U$ if the asymptotic form of the theory is specified.

\section{Asymptotic effective potential}

We shall consider a potential that yields the following expansion around the
AdS vacuum%
\begin{equation}
V=-\frac{3}{\kappa l^{2}}+\frac{1}{2}m^{2}\phi^{2}+\xi\phi^{4}+O(\phi
^{5})\text{ .}%
\end{equation}
We have intentionally omitted a cubic term in the self-interaction, as its
inclusion makes the analysis more complicated due to the existence of
subleading logarithmic branches \cite{Henneaux:2006hk}. Evaluating (\ref{pot})
in the AdS background with coordinates $C=r^{2}$, $A=\frac{r^{2}}{l^{2}%
}+k=B^{-1}$ we get\footnote{For obvious reasons we change the notation from
$x$ to $r$ in this section.}%
\begin{equation}
U=\left(  m^{2}+\frac{2}{l^{2}}\right)  \left(  \frac{r^{2}}{l^{2}}+k\right)
\text{ }. \label{EP}%
\end{equation}

For masses above $m^{2}=-2l^{-2}$ the potential $U$ (\ref{pot}) is
asymptotically positive and divergent. For masses below this one, the
potential $U$ is asymptotically unbounded from below \cite{Hertog:2004bb}.
However, when $m^{2}\geq m_{BF}^{2}$ and for suitable boundary conditions the
spectrum can be positive. The case $m^{2}=-2l^{-2}$ is the limiting case and
we shall focus on it from now on. As it has been worked out in detail in
\cite{Hertog:2004dr, Henneaux:2006hk}, in this case the scalar field fall-off
is
\begin{equation}
\phi=\frac{\alpha}{r}+\frac{\beta}{r^{2}}+O(r^{-3})\text{ }, \label{phi1}%
\end{equation}
where $\alpha$ and $\beta$ denote two functions of the other coordinates. The
relevant fall-off of the metric is
\begin{align}
-g_{tt}  &  =\frac{r^{2}}{l^{2}}+k+O(r^{-1})\text{ },\label{BCA}\\
g_{mn}  &  =r^{2}h_{mn}+O(r^{-1})\text{ },\label{BCB}\\
g_{rr}  &  =\frac{l^{2}}{r^{2}}-\frac{(l^{4}k+\kappa\alpha^{2}l^{2}/2)}{r^{4}%
}+O(r^{-5})\text{ }, \label{BCC}%
\end{align}
Here, $h_{mn}(x^{m})$ is the two-dimensional metric associated to the sphere,
plane or locally hyperbolic space of constant Ricci scalar $2k$, $d\Sigma_{k}%
$. The asymptotic value of the potential is
\begin{equation}
U_{0}:=\lim_{r\rightarrow\infty}U=\frac{3}{2}\frac{\alpha^{2}\left(  8\xi
l^{2}+\kappa\right)  }{l^{4}}\text{ .} \label{AF}%
\end{equation}
Hence, we see that the quartic term governs the asymptotic value of $U$ when
the slow branch is on. As we have just discussed the effective potential for
massless perturbations, $m^{2}=-2l^{-2}$, vanishes in local AdS spacetime
(\ref{EP}). However, the same effective potential evaluated in asymptotically
local AdS spacetime is not asymptotically zero.

As is standard, the tortoise coordinate (\ref{tor}) goes to $-\infty$ at the
black hole horizon and can be set to be zero at the AdS boundary. Hence, the
relevant Schr\"{o}dinger operator has non-trivial self-adjoint extensions and
stability becomes subtler.

\section{A necessary condition for stability}

As the subject of self-adjoint extensions is not so well-known we would like
to start this section with a short heuristic discussion. Let us consider the
operator $H=-\partial_{z}^{2}$. If $z$ belongs to the half real line,
$z\in\left]  -\infty,0\right]  ,$ then $H$ has a square integrable
eigenfunction with a negative eigenvalue,
\begin{equation}
\psi=\exp\left(  \lambda z\right)  \text{ }, \label{bs}%
\end{equation}%
\begin{equation}
\psi=\exp\left(  \lambda z\right)  \Longrightarrow H\psi=E\psi=-\lambda
^{2}\psi\text{ .}%
\end{equation}
with $\lambda>0$. The existence of this eigenvalue is related to the boundary
conditions that the eigenfunction has. It satisfies mixed boundary conditions%
\begin{equation}
b\psi(0)=a\psi^{\prime}\left(  0\right)  \text{ ,} \label{con 2}%
\end{equation}
with
\begin{equation}
\lambda=\frac{b}{a}\text{ .}%
\end{equation}
$a=0$ is the Dirichlet boundary condition, $b=0$ is the Neumann boundary
condition and any other combination are Robin boundary conditions. In this
case, $H$ admits a one-parameter family of self adjoint extensions,
parameterized by $\lambda$. It is easy to see that the negative eigenvalue
only exists for Robin boundary conditions. Furthermore, $\psi$ is square
integrable only when the ratio of $a$ and $b$ is positive. Still, when
$\lambda<0$ the self-adjoint extensions exists but the negative energy state
is excluded from the spectrum.

Let us consider now the operator $\mathcal{H}=-\partial_{z}^{2}+U$, defined on
the half real line as before and let us consider a, square integrable, energy
eigenstate $\psi$. It satisfies%
\begin{equation}
\mathcal{H}\psi=E\psi\Longrightarrow\psi\mathcal{H}\psi=E\psi^{2}\text{ ,}
\label{I1}%
\end{equation}
Integrating by parts (\ref{I1}) yields%
\begin{equation}
-\left.  \psi\partial_{z}\psi\right\vert _{-\infty}^{0}+\int_{-\infty}%
^{0}\left[  \left(  \partial_{z}\psi\right)  ^{2}+U\psi^{2}\right]
dz=E\int_{-\infty}^{0}\psi^{2}dz\text{ .} \label{I2}%
\end{equation}
Using the boundary conditions and the fact that $\psi$ is a bound state,
$\psi(-\infty)=0,$ (\ref{I2}) yields%
\begin{equation}
-\lambda\psi\left(  0\right)  ^{2}+\int_{-\infty}^{0}\left[  \left(
\partial_{z}\psi\right)  ^{2}+U\psi^{2}\right]  dz=E\int_{-\infty}^{0}\psi
^{2}dz\text{ .} \label{I3}%
\end{equation}
As in the case where the potential vanishes we see that $U\geq0$ and
$\lambda\leq0$ automatically excludes negative energies. Moreover, whenever
$U\geq0$, the LHS integral in (\ref{I3}) seems to suggest that it is still
possible to have a positive spectrum for $\lambda>0$; introducing a
competition between the potential, $U$, and the self-adjoint extension
parameter $\lambda$.

Indeed, let us suppose that the spectrum of $\mathcal{H}$ is positive with
$U\geq0$ and $\lambda>0$. As is well-known, given the zero-energy
Schr\"{o}dinger operator%
\begin{equation}
\mathcal{H}\chi=0\text{ .} \label{BC}%
\end{equation}
the number of nodes of $\chi$, count the number of bound states of
$\mathcal{H}$. The linearity of the problem allows to set the boundary
conditions in the simplified form
\begin{equation}
\chi(0)=1\qquad,\qquad\chi^{\prime}(0)=\lambda\text{ .}%
\end{equation}
We shall assume that $\chi$ has no node. Hence, there must be a $z_{0}$, where
$\chi^{\prime}(z_{0})=0$. Notice that $U>0\Longleftrightarrow\chi
^{\prime\prime}>0\Longrightarrow$ $z_{0}$ is a local minimum of $\chi$. The
integral of (\ref{BC}) between $z_{0}$ and $0$ yields%

\begin{equation}
\lambda=\int_{z_{0}}^{0}U\chi dz\text{ .}%
\end{equation}
A bound on $\lambda$ then follows from
\begin{equation}
\int_{z_{0}}^{0}U\chi dz<\chi(0)\int_{z_{0}}^{0}Udz=\int_{z_{0}}^{0}%
Udz<\int_{-\infty}^{0}Udz\text{ .}%
\end{equation}
A necessary condition for stability is then%
\begin{equation}
\lambda<\int_{-\infty}^{0}Udz\text{ .} \label{SO2}%
\end{equation}
In terms of the original perturbation
\begin{equation}
\phi_{1}=\frac{\alpha_{1}}{r}+\frac{\beta_{1}}{r^{2}}+O(r^{-3})\text{ ,}
\label{con1}%
\end{equation}
$\psi(0)=\alpha_{1}$, $\psi^{\prime}(0)=-\beta_{1}l^{-2}$\ we get%
\begin{equation}
\lambda=-\frac{\beta_{1}}{\alpha_{1}l^{2}}\text{ .} \label{SO}%
\end{equation}
It is now clear that not all the boundary conditions allow for linear
stability due to the existence of non-trivial self-adjoint extensions. Indeed,
if the original scalar-tensor theory has either Dirichlet or Neumann boundary
conditions the equation for the perturbation does not admit self-adjoint
extensions. When the prescribed dynamics is in terms of Robin boundary
conditions, the boundary condition of the perturbation can be read-off from
the following equality
\begin{equation}
\phi=\frac{\alpha}{r}+\frac{\beta(\alpha)}{r^{2}}+O(r^{-3})=\phi
_{0}(r)+\epsilon\phi_{1}(r,t)\text{ .}%
\end{equation}
Indeed, using that the background configuration is such that%

\begin{equation}
\phi_{0}=\frac{\alpha_{0}}{r}+\frac{\beta(\alpha_{0})}{r^{2}}+O(r^{-3})\text{
}, \label{SC}%
\end{equation}
it follows that the perturbation satisfy%
\begin{equation}
\phi_{1}=\frac{\alpha_{1}}{r}+\frac{\alpha_{1}\beta^{\prime}(\alpha_{0}%
)}{r^{2}}+O(r^{-3})\text{ }. \label{con4}%
\end{equation}
Thus, comparing (\ref{SO2}), (\ref{con1}), (\ref{SO}) and (\ref{con4}) we find
that our necessary condition for stability, in terms of the boundary condition
of the scalar-tensor theory is%
\begin{equation}
\beta^{\prime}(\alpha_{0})+l^{2}\int_{-\infty}^{0}Udz>0\text{ }. \label{CON2}%
\end{equation}
We could have written the bound in terms of $\int_{z_{0}}^{0}Udz$, however in
practice is much more useful the bound in terms of the full integral.

When the Schr\"{o}dinger operator is defined in the whole real line there is
an estimate of the value of the lowest eigenvalue due to Simon
\cite{Simon:1976un}. It can also be seen as a necessary condition, for the
positivity of the spectrum:
\begin{equation}
\int_{-\infty}^{\infty}Udz>0\text{ .}%
\end{equation}

Our result may be seen as a generalization of it.

\section{Holography and effective potentials}

In the dual field theory, $\alpha$ represents the expectation value of the
operator that condensates in the hairy black hole states. Effective potentials
built from supergravity data are an important tool to investigating the
stability of the equilibrium states under perturbations of the condensate.

Using the standard AdS/CFT dictionary it was proposed in \cite{Hertog:2004ns}
that the effective field theory potential induced by the scalar field is%
\begin{equation}
\mathcal{V}(\alpha)=\int_{0}^{\alpha}\left[  \beta(\alpha)-\beta_{S}%
(\alpha)\right]  d\alpha\text{ ,} \label{EFP}%
\end{equation}
where $\beta(\alpha)$ are the boundary conditions and $\beta_{S}(\alpha)$ is
the soliton line or a regularity condition that can be build as follow. Take a
finite value of the scalar field at the origin and starting from this value
shoot up towards infinity. The fall-off at infinity is fixed by the scalar
field mass. The coefficients of the leading and subleading terms provide a
point in the ($\alpha,\beta$)-plane. Repeat the operation several times and
construct the line $\beta_{S}(\alpha)$. This prescription is such that the
solitons are extremum of the effective field theory potential (\ref{EFP}).

In the well-known $5$-dimensional case when the boundary topology is $R\times
S^{3}$, there is a non-vanishing Casimir energy and the vacuum gravitational
energy of global $AdS_{5}$ can be exactly matched to the vacuum energy of the
large N limit of $\mathcal{N}=4$ super Yang-Mills (see, e.g.,
\cite{Balasubramanian:1999re}). The Hertog-Horowitz prescription also points
out towards the identification of the gravitational energy and the energy of
the field theory dual in designer gravity in the sense that the minimum energy
solution is precisely the hairy soliton that corresponds to the minimum of the
effective potential.

The prescription (\ref{EFP}) is motivated by the fact that at finite
temperature and $\alpha=0$, the gravitational solution is global AdS
with zero mass. Therefore, a natural finite temperature
generalization of (\ref{EFP}) is as follows \cite{Hertog:2005hu}.
Take the Schwarzschild-AdS black at mass $M$. For large enough $M$,
there are boundary conditions, $\beta_{M}(\alpha)$ that ensure the
existence of a hairy black hole with the same mass. The finite
temperature effective field theory potential is:
\begin{equation}
\mathcal{V}_{M}(\alpha)=C\int_{0}^{\alpha}\left[  \beta(\alpha)-\beta
_{M}(\alpha)\right]  d\alpha+M\text{ ,} \label{FT}%
\end{equation}
where $C$ is a positive constant that depends on the theory and allows to
interpret $\mathcal{V}_{M}(\alpha)$ as the exact energy of the gravitational
system. Note that there may be a critical value of the VEV where the hairy
black hole ceases to exists and the line $\beta_{M}(\alpha)$ describe null or
timelike singularities of the same mass. We exclude the singularities from our
analysis. The hairy black holes of mass $M$ are indeed critical points of
(\ref{FT}). The condition for hairy black hole stability is the convexity of
the effective potential
\begin{equation}
\beta^{\prime}(\alpha_{0})-\beta_{M}^{\prime}(\alpha_{0})>0 \label{COND}%
\end{equation}
The most stringent bound that we found in the previous section can be written
as
\begin{equation}
\beta^{\prime}(\alpha_{0})+l^{2}\int_{z_{0}}^{0}Udz>0
\end{equation}
Hence, we conclude that%
\[
-l^{2}\int_{z_{0}}^{0}Udz>\beta_{M}^{\prime}(\alpha_{0})
\]

Note that we are able to obtain a bound on $\beta_{M}^{\prime}(\alpha_{0})$
with our best estimate, given by the LHS integral. The idea is that
(\ref{COND}) represents the exact condition for stability. In the analysis of
the previous section we have found an estimate that bounds its value. We do
not claim that it is a sharp bound. Conversely, this opens the interesting
scenario of providing sharp bounds for Schr\"{o}dinger operators on the
half-line determining the exact effective potential at finite temperature of a
given field theory dual.

\section{Conclusions}

Let us explore in retrospective the linear stability of $AdS_{4}$. The
relevant operator is $H=-\partial_{z}^{2}$. $z$ belongs to the interval,
$z\in\left[  -\frac{\pi}{2}l,0\right]  ,$ where the boundary is at $z=0$. The
regular perturbation with negative energy and generalized boundary conditions
is $\psi\left(  z\right)  =\sinh\left[  \lambda\left(  z+\frac{\pi l}%
{2}\right)  \right]  $. The relevant combination at the boundary is
\begin{equation}
\frac{\psi^{\prime}\left(  0\right)  }{\psi(0)}=\lambda\frac{\cosh\left(
\frac{\pi l}{2}\lambda\right)  }{\sinh\left(  \frac{\pi l}{2}\lambda\right)
}>\frac{2}{\pi l}\text{ .}%
\end{equation}
In terms of the boundary condition of the scalar tensor theory we find
$AdS_{4}$ instabilities for $-\beta^{\prime}(\alpha=0)>\frac{2l}{\pi}$.
Namely, linear vacuum stability is ensured whenever%
\begin{equation}
0\leq\frac{2l}{\pi}+\beta^{\prime}(\alpha=0)\text{ .} \label{Bound}%
\end{equation}
Going back to the discussion in the first paragraph of the introduction, and
with the hindsight of our results, we see that it may very well be possible to
have hairy black holes in theories where the vacuum is linearly stable. In
particular, $AdS$ invariant boundary conditions, $\beta=C\alpha^{2},$ satisfy
trivially the bound (\ref{Bound}). This makes the study of the general
families of hairy black holes at the conformal mass indeed relevant.

We expect that the competition between the boundary condition and the
effective potential of the master equation can be generalized to other cases,
namely when the effective potential has a negative region. It may very well be
possible as well that adequate boundary conditions could render the theory
stable even when the effective potential is everywhere negative. This
understanding is particularly relevant for gauged supergravity.

In four dimensions, there is the well known gauged $\mathcal{N}=8$
supergravity \cite{deWit:1982ig}. The mass of the $70$ scalars,
around the maximally supersymmetric $AdS_{4}$ vacuum, is
$m^{2}=-2l^{-2},$ see for instance Table $2$ in
\cite{DallAgata:2011aa} and references therein, which reflects the
fact that they are effectively massless around this vacuum. In a
remarkable paper \cite{Dall'Agata:2012bb} it has been shown that the
de Wit-Nicolai gauged $\mathcal{N}=8$ supergravity is not unique,
but a single member of a one parameter ($\omega$) family of
supergravities. As discussed in \cite{Tarrio:2013qga}, two of the
single scalar field truncations allow for a non-trivial
$\omega-$deformation. For related work see also \cite{Pang:2015mra}.

The four single-scalar field consistent truncations of the $SL(8,\mathbb{R}%
)/SO(8)$ sector, worked out in detail in \cite{Cvetic:1999xx}, see also Table
$1$ of \cite{Papadimitriou:2006dr}, with a possible $\omega-$deformation, can
be described by the scalar field self-interaction introduced in
\cite{Anabalon:2012ta}. It also provide infinitely many examples that can be
embedded in gauged $\mathcal{N}=1$ supergravity \cite{Anabalon:2013eaa} and as
recently remarked in \cite{Faedo:2015jqa} in gauged $\mathcal{N}=2$
supergravity with a Fayet-Iliopoulos term. We shall apply the methods
discussed here to provide a throughout study of the hairy black hole stability
in these theories in a future work.

\section*{Acknowledgements}
A.A. thanks the enlightening discussions with Simon Ross about
stability, Ioannis Papadimitriou for clarifying several aspects of
the holographic interpretation and the hospitality of Durham
University where this work was boosted. A.A. also thanks the the
hospitality of Wellington Galleas at DESY and Ji\v{r}\'{\i}
Bi\v{c}\'{a}k at Charles University where this work was continued
and completed (supported  from the Grant No. 14-37086G, Albert
Einstein Center). We would like to thank Cristi\'{a}n Martinez and
Ra\'{u}l Rojas for interesting discussions and collaboration on
related projects. Research of A. A. is supported in part by the
Fondecyt Grants No 11121187 and 1141073, and by the Newton-Picarte
Grants DPI20140053 and DPI20140115. Research of D.A. has been
partially funded by the Fondecyt grants 1120446 and by the
Newton-Picarte Grant DPI20140115. J.O. is supported by FONDECYT
grant 1141073 and by the Newton-Picarte Grant DPI20140053.

\end{document}